\documentstyle[12pt]{article}
\newcommand{\bea}{\begin{eqnarray}}
\newcommand{\eea}{\end{eqnarray}}
\newcommand{\beq}{\begin{equation}}
\newcommand{\eeq}{\end{equation}}
\newcommand{\bay}{\begin{array}}
\newcommand{\eay}{\end{array}}
\begin{document}


\begin{center}
{\Large \bf Order-$\alpha_s^3$ determination of the
 strange quark mass}\\[1cm]

{\bf K.G. Chetyrkin$^1$, D. Pirjol$^2$ and K. Schilcher$^3$}\\[.5cm]
$^1$Institute for Nuclear Research of the Russian Academy of Sciences,\\
60$^{th}$ October Anniversary Prospect 7a,\\
117312 Moscow, Russia\\[.75cm]
$^2$Department of Physics, Technion - Israel Institute of Technology,\\
32000 Haifa, Israel\\[.75cm]
$^3$Institut f\"{u}r Physik,
Johannes Gutenberg-Universit\"{a}t\\
Staudingerweg 7, D-55099 Mainz, Germany\\
\end{center}
\vspace{0.5cm}
\begin{abstract}
\noindent
We present a QCD sum rule calculation of the strange-quark mass including
four-loop QCD corrections to the correlator of scalar currents.
We obtain $\bar m_s(1$ GeV$)=205.5\pm 19.1$ MeV.
\end{abstract}
\newpage
\section*{1.Introduction}

   A precise determination of the values of the light quark masses is of
crucial practical importance for testing in an accurate way the predictions
of the Standard Model. In particular, the knowledge of the strange quark mass 
is relevant for a better understanding of the low-energy phenomenology of
QCD and for a precise prediction of the CP-violating parameter $\epsilon'/
\epsilon$ in the framework of the Standard Model \cite{eps1,eps2,eps3,eps4}.

   The ratios of the light quark masses can be determined
in a model-independent way with the help of chiral-perturbation methods
\cite{ChPT}. On the other hand, in order to obtain their absolute values, one
has to resort either to the method of QCD sum rules \cite{QCDSR} or to
lattice QCD \cite{lattice,lattice2}. While no less fundamental than the
first type of predictions, the
latter ones have suffered in the past from larger uncertainties. 
In \cite{ChDo} a QCD sum rule calculation of the strange
quark mass in the scalar channel has been presented, which employed N$^2$LO
(${\cal O} (\alpha_s^2)$) results for the correlator of two scalar
strangeness-changing currents in perturbation theory (see also \cite{JaMu} for
a similar calculation) (for earlier calculations, see
\cite{Nar1,Nar2,NPRT,CL,DvGP}). 

      In the meantime the N$^3$LO
(${\cal O}(\alpha_s^3)$) correction to this correlator has become also known
\cite{Chet}. The perturbative contribution dominates the sum rule
\cite{ChDo,JaMu},
so one naturally expects the N$^3$LO correction to alter these results in a
significant way. This expectation is also supported by a simple
estimate of these corrections \cite{JaMu} which shows that their omission is
likely to constitute the main source of errors in the calculations of
\cite{ChDo} and \cite{JaMu}.

   Since the N$^3$LO correction is now known, we are in a position to 
present a reevaluation of the strange quark mass computation given in
\cite{ChDo}.

\section*{2. Four-loop contributions to the scalar correlator}

     The QCD sum rule used in this paper is based on the correlator of two 
scalar currents
\bea\label{psi}
\psi (Q^2,\alpha_s,m_s,\mu) = i\int\mbox{d}x e^{iqx}
\langle 0|\mbox{T} J(x) J^\dagger(0) |0\rangle
\eea
where $J = \partial_\alpha\bar s\gamma^\alpha u = i(m_s-m_u)\bar su$,
$Q^2=-q^2$.  It will be more convenient to work with the second derivative of
$\psi(Q^2)$, $\psi''(Q^2) = \mbox{d}^2/\mbox{d}(Q^2)^2$, which satisfies
a homogeneous renormalization-group equation $\mu\frac{\mbox{d}}{\mbox{d}\mu}
\psi''(Q^2) = 0$.

 We will write $\psi''(Q)$ as $\psi_P''(Q)+\psi_{NP}''(Q)$, with $\psi_P''
(Q)$ is the perturbative part and
$\psi_{NP}''(Q)$ contains the vacuum expectation values of the higher 
dimension operators.  For the perturbative part one obtains the following
result:
\bea
\psi_P''(Q) &=& \frac{6(m_s-m_u)^2}{(4\pi)^2Q^2}\left\{
1 + \frac{11\alpha_s}{3\pi} + \frac{\alpha_s^2}{\pi^2}\left(
\frac{5071}{144}-\frac{35}{2}\zeta(3)\right)\right.\nonumber\\
& &\left.\hspace{3cm} +\frac{\alpha_s^3}{\pi^3}
\left(-\frac{4781}{9}+\frac16 a_1+\frac{475}{4}\zeta(3)\right)\right.
\nonumber\\
&+&\left.\log\frac{Q^2}{\mu^2}\left[
-2\frac{\alpha_s}{\pi} - \frac{139\alpha_s^2}{6\pi^2} +
\frac{\alpha_s^3}{\pi^3}\left(-\frac{2720}{9}+\frac{475}{4}\zeta(3)\right)
\right]\right.\label{psi''}\\
&+&\left.\log^2\frac{Q^2}{\mu^2}\left[\frac{17\alpha_s^2}{4\pi^2}
 + \frac{695\alpha_s^3}{8\pi^3}\right]
- \frac{221\alpha_s^3}{24\pi^3}\log^3\frac{Q^2}{\mu^2}\right\}
\nonumber\\
& &\hspace{-1cm} -\frac{12(m_s-m_u)^2m_s^2}{(4\pi)^2Q^4}\left\{1 +
\frac{28\alpha_s}{3\pi}
+ \frac{\alpha_s^2}{\pi^2}\left(\frac{8557}{72}-\frac{77}{3}\zeta(3)\right)
\right.\nonumber\\
&-&\left.\log\frac{Q^2}{\mu^2}\left[4\frac{\alpha_s}{\pi} +
\frac{147\alpha_s^2}{2\pi^2}\right]
+ \frac{25\alpha_s^2}{2\pi^2}\log^2\frac{Q^2}{\mu^2}\right\}\,,\nonumber\\
a_1 &=& \frac{4748953}{864}-\frac{\pi^4}{6}-\frac{91519\zeta(3)}{36}+
\frac{715\zeta(5)}{2} \simeq 2795.0778\,.\nonumber
\eea

The terms of order $\alpha_s^3$ in the $O(m_q^2)$
part of (\ref{psi''}) have been extracted from the recent
four-loop calculation of \cite{Chet}.  The terms of order
$\alpha_s^2$ in the $O(m_q^4)$ part of (\ref{psi''}) can be
found in \cite{Chetprep}.

The exact value of $a_1$ agrees well with an estimate \cite{JaMu} of the
same quantity\footnote{The corresponding constant in \cite{JaMu} is called
$c_{31}=a_1/6$.} based on the assumption of a continued geometric growth
of the perturbative series for $\psi''(Q^2)$, which gave $a_1=2660$.
 
We have neglected the light quark mass $m_u$, except in the overall factors.
The renormalized parameters $\alpha_s$ and $m_s,m_u$ are taken at
the scale $\mu$. Their $\mu$-dependence should cancel against that of the
$\log\,\mu$ factors in (\ref{psi''}) so that $\psi''(Q)$ is
$\mu$-independent.

   As for the nonperturbative contributions, we keep
only the dimension-4 operators. These are given, together with their 
renormalization-group properties and the values of the coefficient
functions $c_i$ to next-to-leading order, in \cite{ChDo,JaMu} where 
the references to the original  calculations can also
be found. We quote here only the final result for $\psi_{NP}''(Q)$:
\bea
\psi_{NP}''(Q) &=& \frac{(m_s-m_u)^2}{Q^6}\left\{2\langle m_s\bar uu\rangle_0
\left(1 + \frac{\alpha_s}{\pi}(\frac{23}{3}-2\log\frac{Q^2}{\mu^2})\right)
\right.\nonumber\\
& &\left.\hspace{-1cm} -\frac19 I_G\left(1 + \frac{\alpha_s}{\pi}
(\frac{121}{18} - 2\log\frac{Q^2}{\mu^2})\right)
+ I_s\left(1+\frac{\alpha_s}{\pi}(\frac{64}{9}-2\log\frac{Q^2}
{\mu^2})\right) \right.\label{NP}\\
& &\left.\hspace{-1cm} -\frac{3}{7\pi^2}m_s^4\left(\frac{\pi}{\alpha_s}+
\frac{155}{24} - \frac{15}{4}\log\frac{Q^2}{\mu^2} \right)\right\}\,.\nonumber
\eea
$I_s$ and $I_G$ are the vacuum expectation values of the two RG-invariant
combinations of dimension 4, which are given for $n_f=3$ and to the order we
are working, by
\bea\label{Is}
I_s &=& m_s\langle \bar ss\rangle_0 + \frac{3}{7\pi^2}m_s^4
\left(\frac{\pi}{\alpha_s} - \frac{53}{24}\right)\\
I_G &=& -\frac94\langle\frac{\alpha_s}{\pi}G^2\rangle_0\left(1+
\frac{16}{9}\frac{\alpha_s}{\pi}\right) + \frac{4\alpha_s}{\pi}
\left(1+\frac{91}{24}\frac{\alpha_s}{\pi}\right) m_s\langle\bar ss\rangle_0
+\frac{3}{4\pi^2}\left(1+\frac43\frac{\alpha_s}{\pi}\right)m_s^4\,.\nonumber
\\\label{IG}
\eea

\section*{4.The sum rule}

   To enhance the contribution of the low-lying states, one applies a Borel
transform to the both sides of the dispersion relation used to define the
sum rule \cite{ChDo}.
The effect is to transform the power-suppression of the states
with a large invariant mass into an exponential one,
controlled by the Borel parameter $M^2$:
\bea\label{numint}
\hat L[\psi''(Q^2)] = \frac{1}{M^6}\frac{1}{\pi}\int_0^\infty\mbox{d}t
e^{-t/M^2}\mbox{Im}\,\psi(t)\,.
\eea
The Borel transform of the l.h.s. can be computed from (\ref{psi''},\ref{NP})
and is given by
\bea
\hat L[\psi_P''(Q)] &=& \frac{6(m_s-m_u)^2}{(4\pi)^2M^2}\left\{
1 + \frac{\alpha_s}{\pi}(\frac{11}{3}-2\psi(1)) + \frac{\alpha_s^2}{\pi^2}
\left(\frac{5071}{144}-\frac{35}{2}\zeta(3) + \frac{17}{4}\psi^2(1)
\right.\right.\nonumber\\
& &\left.\left.\hspace{-2cm} -
\frac{139}{6}\psi(1) - \frac{17}{24}\pi^2\right)
+\frac{\alpha_s^3}{\pi^3}
\left(-\frac{4781}{9}+\frac16 a_1+\frac{475}{4}\zeta(3)\psi(1)
+ \frac{823}{6}\zeta(3)\right.\right.\nonumber\\
& &\left.\left.\hspace{-2cm} - \frac{221}{24}\psi^3(1)
 + \frac{695}{8}\psi^2(1)
+ \frac{221}{48}\psi(1)\pi^2 - \frac{2720}{9}\psi(1) - \frac{695}{48}\pi^2
\right) + \log\frac{M^2}{\mu^2}\left[ -2\frac{\alpha_s}{\pi}
\right.\right.\nonumber\\
& &\left.\left.\hspace{-2cm}
 + \frac{\alpha_s^2}{\pi^2}\left(-\frac{139}{6}
+\frac{17}{2}\psi(1)\right) +
\frac{\alpha_s^3}{\pi^3}\left(-\frac{2720}{9}+\frac{475}{4}\zeta(3)
- \frac{221}{8}\psi^2(1) + \frac{695}{4}\psi(1)
\right.\right.\right.\nonumber\\
& &\left.\left.\left.\hspace{-2cm} + \frac{221}{48}\pi^2 \right)
\right] + \log^2\frac{M^2}{\mu^2}\left[\frac{17\alpha_s^2}{4\pi^2}
 + \frac{\alpha_s^3}{\pi^3}(\frac{695}{8} - \frac{221}{8}\psi(1))\right]
- \log^3\frac{M^2}{\mu^2}\cdot\frac{221\alpha_s^3}{24\pi^3}\right\}
\label{psi''Borel}\\
& &\hspace{-1cm} -\frac{12(m_s-m_u)^2m_s^2}{(4\pi)^2M^4}\left\{1 +
\frac{\alpha_s}{\pi}\left(\frac{16}{3}-4\psi(1)\right)
+ \frac{\alpha_s^2}{\pi^2}\left(\frac{5065}{72}-\frac{25}{12}\pi^2
- \frac{97}{2}\psi(1)\right.\right.\nonumber\\
& &\left.\left.\hspace{-2cm} + \frac{25}{2}\psi^2(1)\right)
-\log\frac{M^2}{\mu^2}\left[4\frac{\alpha_s}{\pi} +
\frac{\alpha_s^2}{\pi^2}\left(\frac{97}{2}-25\psi(1)\right)\right]
+ \log^2\frac{M^2}{\mu^2}\cdot \frac{25\alpha_s^2}{2\pi^2}\right\}
\nonumber
\eea
and respectively
\bea
\hat L[\psi_{NP}''(Q)] &=&
\frac{(m_s-m_u)^2}{2M^6}\left\{2\langle m_s\bar uu\rangle_0
\left(1 + \frac{\alpha_s}{\pi}(\frac{14}{3}-2\psi(1)-2\log\frac{M^2}{\mu^2})
\right)\right.\label{NPBorel}\\
& &\left.\hspace{-2cm} -\frac19 I_G\left(1 + \frac{\alpha_s}{\pi}(
\frac{67}{18} - 2\psi(1)-2\log\frac{M^2}{\mu^2})\right)
+ I_s\left(1+\frac{\alpha_s}{\pi}(\frac{37}{9}-2\psi(1)-2\log\frac{M^2}
{\mu^2})\right) \right.\nonumber\\
& &\left.\hspace{-2cm} -\frac{3}{7\pi^2}m_s^4\left(\frac{\pi}{\alpha_s}+
\frac{5}{6} - \frac{15}{4}\psi(1)-\frac{15}{4}\log\frac{M^2}{\mu^2}
\right)\right\}\,.\nonumber
\eea
The numerical constants entering these expressions have the values
$\psi(1)=-\gamma_E=-0.577$ and $\zeta(3)=1.202$. 

   At this point the usual procedure is to take advantage of the
$\mu$-independence of $\hat L[\psi''(Q)]$ and the fact that the operation
of Borel 
transformation does not act on $\mu$) and choose $\mu=M$ \cite{CDKS}. This
``renormalization-group improvement'' effectively shifts the logs of
$M^2/\mu^2$ into the renormalized parameters $\alpha_s(M)$ and $m_s(M)$.
To the order we are working and for $n_f=3$, these are given by
\bea\label{as}
\frac{\alpha_s(M)}{\pi} &=& \frac{4}{9}\frac{1}{L} - \frac{256}{729}
\frac{LL}{L^2} + \frac{1}{L^3}\left( \frac{6794}{59049} -
\frac{16384}{59049}LL + \frac{16384}{59049}LL^2\right)\\
m_s(M) &=& \frac{\hat m_s}{(\frac12 L)^{4/9}}\left\{
1 + \frac{290}{729}\frac{1}{L} - \frac{256}{729}\frac{LL}{L} +
\left( \frac{550435}{1062882}-\frac{80}{729}\zeta(3)\right)\frac{1}{L^2}
\right.\nonumber\\
& &\hspace{-2cm}\left.-\frac{388736}{531441}\frac{LL}{L^2} +
\frac{106496}{531441}\frac{LL^2}{L^2}\right.\label{ms}\\
& &\hspace{-2cm}\left. +
\left(\frac{2121723161}{2324522934}+
\frac{8}{6561}\pi^4-\frac{119840}{531441}\zeta(3)-\frac{8000}{59049}\zeta(5)
\right)\frac{1}{L^3} \right.\nonumber\\
& &\hspace{-2cm}\left.+\left(-\frac{611418176}{387420489}+
\frac{112640}{531441}\zeta(3)
\right)\frac{LL}{L^3} + \frac{335011840}{387420489}\frac{LL^2}{L^3} -
\frac{149946368}{1162261467}\frac{LL^3}{L^3}\right\}\,.\nonumber
\eea
We have denoted here $L=\log(M^2/\Lambda^2_{QCD})$ and $LL=\log\,L$.
We have used in (\ref{ms}) the recently calculated exact values
of the 4-loop beta function \cite{vRVL}
\bea\label{beta4}
\beta_4(n_f=3)=-\frac{140599}{2304} - \frac{445}{16}\zeta(3)\simeq
-94.456\,
\eea
and of the 4-loop mass anomalous dimension \cite{Chetgm,vRVLgm}
\bea\label{gamma4}
\gamma_4(n_f=3)=\frac{2977517}{20736}+\frac{3}{16}\pi^4 - 
\frac{9295}{216}\zeta(3) - \frac{125}{6}\zeta(5)\simeq
88.525817\,
\eea


The exact result (\ref{gamma4}) is close to an estimate based
on the assumption of a geometrical growth
for the coefficients of the $\gamma_m$ anomalous dimension
\bea\label{gamma4est}
\gamma_4(n_f=3)=\frac{\gamma_3^2(n_f=3)}{\gamma_2(n_f=3)}
\simeq 81.368\,,
\eea
with \cite{gamma,Larin:massQCD}
\bea\label{gamma3}
\gamma_1(n_f=3) = 2\,,\qquad \gamma_2(n_f=3) = \frac{91}{12}\,,
\qquad \gamma_3(n_f=3) = \frac{8885}{288}-5\zeta(3)\,.
\eea

%
$$ \vbox{\offinterlineskip
\def\tablerule{\noalign{\hrule}}
\halign {\vrule#& \strut#&
\ \hfil#\hfil& \vrule\,\vrule#&
\ \hfil#\hfil& \vrule#&
\ \hfil#\hfil& \vrule#&
\ \hfil#\hfil& \vrule\,\vrule#&
\ \hfil#\hfil& \vrule#&
\ \hfil#\hfil& \vrule#&
\ \hfil#\hfil\ & \vrule# \cr
\tablerule
height10pt && \omit && \omit && \omit && \omit && \omit
&& \omit && \omit  &\cr
&&  && \,\, $1/L$ \,\,
&& \,\, $1/L^2$ \,\, && \,\, $1/L^3$ \,\, 
&& \,\, $m_s^{(2)},\alpha_s^{(1)}$ \,\, 
&& \,\, $m_s^{(3)},\alpha_s^{(2)}$ \,\, 
&& \,\, $m_s^{(4)},\alpha_s^{(3)}$ \,\, & \cr
height10pt && \omit && \omit && \omit && \omit && \omit
&& \omit && \omit  &\cr
\tablerule
height10pt && \omit && \omit && \omit && \omit && \omit
&& \omit && \omit &\cr
&& $m_s^2$ && 0.6942 && 0.6673 && 0.6638 && 0.6942 && 0.6675 && 0.6641 &\cr
height10pt && \omit && \omit && \omit && \omit && \omit
&& \omit && \omit &\cr
&& $m_s^2(\frac{\alpha_s}{\pi})$ && 0.1011 && 0.0725 && 0.0720 
&& 0.1017 && 0.0695 && 0.0727 &\cr
height10pt && \omit && \omit && \omit && \omit && \omit
&& \omit && \omit &\cr
&& $m_s^2(\frac{\alpha_s}{\pi})^2$ && --- && 0.0148 && 0.0063
&& --- && 0.0072 && 0.0080 &\cr
height10pt && \omit && \omit && \omit && \omit && \omit
&& \omit && \omit &\cr
&& $m_s^2(\frac{\alpha_s}{\pi})^3$ && --- && --- && 0.0022
&& --- && --- && 0.0009 &\cr
height10pt && \omit && \omit && \omit && \omit && \omit
&& \omit && \omit &\cr
\tablerule}} $$
{\bf Table 1.}
Values of $m_s^2(M^2)(\frac{\alpha_s(M^2)}{\pi})^i$ at $M^2=3$ GeV$^2$
($\Lambda_{\overline{MS}}^{n_f=3}=380$ MeV) used for the discussion of
the validity of the truncation approximation in the text.\\[0.3cm]

The usual approach followed in the numerical evaluation of the sum rule
\cite{ChDo,JaMu} has been to expand the Borel transforms (\ref{psi''Borel}) 
and (\ref{NPBorel}) in powers of $1/L$ and truncate the resulting 
expressions to a given order in this parameter. There are certain
errors inherent to this procedure which actually turn out to be important
in practice.
This can be seen at order $1/L^2$ 
by examining the structure of the Borel transform
of the leading term in (\ref{psi''Borel})
\bea\label{T1}
\hat T_1(M^2) = \frac{6m_s^2(M^2)}{(4\pi)^2M^2}
\left(1 + \frac{\alpha_s(M^2)}{\pi}c_1 + \left(\frac{\alpha_s(M^2)}{\pi}
\right)^2 c_2 + \left(\frac{\alpha_s(M^2)}{\pi}
\right)^3 c_3 + \cdots\right)\,,
\eea
where we neglected the light quark mass.
The first three coefficients $c_i$ have the values $c_1=4.8211, c_2=21.9765,
c_3= 53.1421$.

We tabulated in Table 1 the values of the quantities $m_s^2(M^2)
(\alpha_s(M^2)/\pi)^i$ (with $\hat m_s=1$) for $i=0-3$ using two different
approximations at a
typical value of the Borel parameter $M^2=3$ GeV$^2$. The first three
columns show the values of these parameters computed by expanding in
powers of the small parameter $1/L$ up to the shown order. This 
approximation has been commonly used in the previous literature (for
example the results in \cite{ChDo,JaMu} have been obtained using a
similar expansion to order $1/L^2$). 

 From Table 1 one can see that truncating $m_s^2(\alpha_s/\pi)^2$ to order
$1/L^2$ results in an error of the order of 100\%, as the $1/L^3$ correction
to its value is comparable with the $1/L^2$ term itself. The result
of truncating to order $1/L^2$ is to overestimate the ${\cal O}(\alpha_s^2)$
correction to the sum rule by a factor of 2. Neglecting this fact could result
in the
paradoxical consequence that adding the ${\cal O}(\alpha_s^3)$ correction
decreases the perturbative contribution $\hat T_1(M^2)$ to the sum rule,
although each separate term in the $\alpha_s$ expansion is positive!

The last three columns of Table 1 show the untruncated values for these
parameters (e.g. $m_s^{(i)},\alpha_s^{(j)}$ means that the full $i$-loop
expression for the running mass and the $j$-loop one for the running
coupling have been used). One can see that these quantities are more
stable when going from one order in perturbation theory to another than
the truncated quantities. They can be therefore expected to give a closer
estimate of the true size of each correction term and will be used in
our numerical analysis below.

Another difference from the treatment followed in \cite{ChDo,JaMu}
will be the use of the 4-loop formulas for the running parameters
(\ref{as},\ref{ms}) in all our expressions 
(\ref{psi''Borel}), (\ref{NPBorel}) and (\ref{ImPsi}).
We recall that the truncation approach (working to a finite order 
in $1/L$) employs running parameters at lower orders in
the loop expansion in the power-supressed terms. The
numerical differences between these two approaches are not
significant. However, the former one is physically preferable
as it uses as small expansion parameter $\alpha_s(M)$ which is
what is directly measured in practice (as opposed to $1/L$).

The condensates entering the renormalization-group
invariant condensates (\ref{Is}) and (\ref{IG}) are taken, as in \cite{ChDo},
at the reference scale $\mu_0=1$ GeV and have the values $\langle\bar uu
\rangle|_{\mu_0}=-(0.225)^3$ GeV$^3$ and $\langle\frac{\alpha_s}{\pi}
G^2\rangle=0.02-0.06$ GeV$^4$. The amount of SU(3)-breaking in the scalar
condensate $\langle\bar ss\rangle/\langle\bar uu\rangle$ will be varied
between 0.7 and 1. The up quark mass has been taken \cite{PDG} as
$\bar m_u(1$ GeV)=5 MeV.

    The hadronic contribution to the sum rule is expressed below 
$s_0=6-7$ GeV$^2$ in terms of the scalar
form-factor $d(s)$ in $K_{\ell 3}$ decays. One has \cite{NPRT}
\bea
\frac{1}{\pi}\mbox{Im}\,\psi(s) &=& \frac{3}{32\pi^2}|d(s)|^2
\sqrt{\left(1-\frac{(m_K+m_\pi)^2}{s}\right)\left(1-\frac{(m_K-m_\pi)^2}{s}
\right)}\,.
\eea

The values of $s$ appearing in this relation are not accessible in $K_{\ell
3}$ decays, but extend above the $K\pi$-production threshold $s_{th}=
(m_K+m_\pi)^2$. 
We will parametrize the scalar form-factor $d(s)$ in this region by a sum of
Breit-Wigner resonances
corresponding to the two bound states with the quantum numbers of the scalar
current $K_0^*(1430)$ and $K_0^*(1950)$ \cite{CL,JaMu}

\bea\label{BW}
|d(s)|^2 = |d(s_{th})|^2\frac{\displaystyle
\frac{\Gamma_1}{(M_1^2-s)^2+M_1^2\Gamma_1^2(s)} +
\frac{\gamma^2\Gamma_2}{(M_2^2-s)^2+M_2^2\Gamma_2^2(s)}
}
{\displaystyle
\frac{\Gamma_1}{(M_1^2-(m_K+m_\pi)^2)^2} +
\frac{\gamma^2\Gamma_2}{(M_2^2-(m_K+m_\pi)^2)^2}
}\,.
\eea

The threshold value of the scalar form-factor $d(s_{th})$ has been computed
to order $p^4$ in chiral perturbation theory \cite{ChPT} with the result
$|d(s_{th})|^2= 0.35$ GeV$^2$. The same quantity has 
been recently extracted \cite{JaMu} from data on $s$-wave phase shifts for
$K\pi$ scattering \cite{exp} with a similar result $|d(s_{th})|^2 = 0.33\pm
0.02$ GeV$^2$. The values of the other parameters in (\ref{BW}) are
\cite{JaMu}
$M_1=1423\pm 10$ MeV, $\Gamma_1=268\pm 25$ MeV, $M_2=1945\pm 22$ MeV,
$\Gamma_2=201\pm 86$ MeV, $\gamma=0.5\pm 0.3$. The energy-dependent
widths $\Gamma_i(s)$ are given by
\bea
\Gamma_i(s) = \Gamma_i\sqrt{\frac{\displaystyle
\left(1-\frac{(m_K+m_\pi)^2}{s}\right)\left(1-\frac{(m_K-m_\pi)^2}{s}\right)
}
{\displaystyle
\left(1-\frac{(m_K+m_\pi)^2}{M_i^2}\right)
\left(1-\frac{(m_K-m_\pi)^2}{M_i^2}\right)
}}\,.
\eea
   For the region of large invariant mass of the hadronic states ($s>s_0$),
parton-hadronic duality can be expected to hold to a high degree of
precision. This allows us to take
the spectral density Im $\psi(s)$ to be equal to the imaginary
part of the QCD expression (\ref{psi''}). This is
given by
\bea\label{ImPsi}
\frac{1}{\pi}\mbox{Im}\,\psi(s) &=& \frac{3}{8\pi^2}(m_s-m_u)^2s\left\{1 +
\frac{\alpha_s}{\pi}\left(\frac{17}{3}-2\ln\frac{s}{\mu^2}\right)\right.\\
& &\hspace*{-1.5cm}\left.+\frac{\alpha_s^2}{\pi^2}\left(
-\frac{35\zeta(3)}{2} + \frac{9631}{144}
-\frac{17\pi^2}{12}-\frac{95}{3}\ln\frac{s}{\mu^2}+\frac{17}{4}
\ln^2\frac{s}{\mu^2}\right)\right.\nonumber\\
& &\hspace*{-1.5cm}\left.+ \frac{\alpha_s^3}{\pi^3}\left[
\frac{4748953}{5184}-\frac{\pi^4}{36}
-\frac{91519\zeta(3)}{216}+\frac{715\zeta(5)}{12}-\frac{229\pi^2}{6}
\right.\right.\nonumber\\
& &\left.\left.+
\left(-\frac{4781}{9}+\frac{221\pi^2}{24}+\frac{475\zeta(3)}{4}\right)
\ln\frac{s}{\mu^2} + \frac{229}{2}\ln^2\frac{s}{\mu^2}-
\frac{221}{24}\ln^3\frac{s}{\mu^2}\right]\right\}\nonumber\\
& &\hspace*{-1.5cm}-\frac{3}{4\pi^2}(m_s-m_u)^2m_s^2\left\{1 +
\frac{\alpha_s}{\pi}\left(\frac{16}{3}
-4\ln\frac{s}{\mu^2}\right)\right.\nonumber\\
& &\left. + \frac{\alpha_s^2}{\pi^2}\left(
-\frac{77\zeta(3)}{3}+\frac{5065}{72}-
\frac{25\pi^2}{6}-\frac{97}{2}\ln\frac{s}{\mu^2}+
\frac{25}{2}\ln^2\frac{s}{\mu^2}\right)\right\}\nonumber\\
& &\hspace*{-1.5cm}+ \frac{m_s^2(s)}{s}\left\{ \frac{45}{56\pi^2}m_s^4(s) -
\frac{2\alpha_s(s)}{\pi}\langle m_s\bar uu\rangle_0 +
\frac{\alpha_s(s)}{9\pi}I_G - \frac{\alpha_s(s)}{\pi}I_s\right\}\nonumber
\,.
\eea
The integration of this expression over $t=(s_0,\infty)$ in (\ref{numint})
is performed numerically keeping $\mu$ arbitrary, after which $\mu$ is
set equal to $M^2$. At this point we have again the option of truncating
or not the expression obtained after
expanding $m_s(M),\alpha_s(M)$ according to (\ref{as},\ref{ms})
to a given power of $1/\ln(\Lambda^2/M^2)$.
In accordance with the treatment of the theoretical side of the sum rule
discussed above, we choose not to truncate the integral of the perturbative
discontinuity either.

\section*{5. Results and discussion}

  In Fig.1 are presented plots of the invariant mass $\hat m_s$
and of the running mass at the scale 1 GeV as a function
of the Borel parameter $M^2$ for different values
of the continuum threshold $s_0$ and the central value of the QCD scale
$\Lambda^{n_f=3}_{\overline{MS}} = 380$ MeV \cite{Be}.  We extract our
results from
the region in $M^2$ corresponding to the stability interval $M^2=2-9$ GeV$^2$,
obtaining in this way $\hat m_s = 172-191$ MeV respectively
$m_s(1$ GeV)=191-213 MeV.
 The error arises mainly
from the $s_0$ and $M^2$ dependence, the errors due to the condensates being
negligible, under 1-2\%.

The effect on $\hat m_s$ of changing $\Lambda^{n_f=3}_{\overline{MS}}$ between
the limits 280-480 MeV is shown in Fig.2.
The continuum threshold has been chosen such that optimal stability is
obtained for each value of $\Lambda^{n_f=3}_{\overline{MS}}$.
For $\Lambda^{n_f=3}_{\overline{MS}} = 280,380,480$ MeV we find
$s_0=5.0,6.0$ and 6.9 GeV$^2$. The corresponding values for the
invariant mass are $\hat m_s=231-232$ MeV, 181-182 MeV and 140-147 MeV.
This rather large spread of values is considerably reduced for the running 
mass at the scale 1 GeV, for which we obtain $m_s(1$ GeV) = 209-210 MeV,
201-202 MeV and 211-221 MeV.

As one can see from Fig.2 the larger values of $m_s(1$ GeV) arises from
including
the large value of the QCD scale $\Lambda^{n_f=3}_{\overline{MS}} = 480$
MeV. If this
curve is eliminated the following results are obtained:
\bea\label{res1}
\hat m_s = 181-232 \mbox{ MeV}\,,\qquad \bar m_s(1 GeV) = 201-210
\mbox{ MeV}\,.
\eea
A similar observation has been made in \cite{Sh} in the context of the
QCD sum rule for the $\rho$ meson width, where even lower values for
$\Lambda^{n_f=3}_{\overline{MS}}$ are advocated, of the order of 220 MeV.
We will adopt therefore in the following (\ref{res1}) as our result 
incorporating the theoretical errors arising from varying
$\Lambda^{n_f=3}_{\overline{MS}}=280-380$ MeV.

   These results are significantly higher than the ${\cal O}(\alpha_s^2)$
results of \cite{ChDo,JaMu}, so that an explanation for this difference is
necessary. As mentioned already (see the discussion surrounding Eq.(16))
the approach used is this paper differs from that of \cite{ChDo,JaMu} in
that the Borel transform is not expanded in powers of $1/\ln(\Lambda^2/M^2)$
but all orders in this parameter are kept. As an effect the leading term
in the perturbative contribution to the sum rule (of ${\cal O}(m_s^2/M^2)$)
is smaller than in \cite{ChDo,JaMu}, even after including the 4-loop
contribution, by about 8\%. This results in an increase in the invariant
mass by 4\%, respectively 7-9 MeV. Another effect which pushes the result
to the high side is the increase of the ${\cal O}(m_s^4/M^4)$ term, when
adding the ${\cal O}(\alpha_s^2)$ contribution. Since this term 
contributes with a negative sign to the theoretical side of the sum rule,
it results also in a small increase of 1-2 MeV in the final result.

For purposes of comparation we give also the results obtained
if both sides of the sum rule had been truncated
to order $1/L^3$ (in the leading terms of ${\cal O}(m_s^2/M^2)$).
For $\Lambda^{n_f=3}_{\overline{MS}} = 380$ MeV the best stability is
obtained for $s_0=5.7$ GeV$^2$ and the results for the strange quark
mass are $\hat m_s=175-176$ MeV, respectively\\ $m_s(1$ GeV)=$192-193$ MeV.
Changing the continuum threshold $s_0$ by $\pm$0.5 GeV$^2$ about this
value gives the
broader range of values $\hat m_s=167-184$ MeV, $m_s(1$ GeV)=$183-202$ MeV.
Choosing $\Lambda^{n_f=3}_{\overline{MS}}=280,480$ MeV gives
(for $s_0=4.8,6.7$ GeV$^2$) the results 
$\hat m_s=225-227, 135-139$ MeV, respectively 
$m_s(1$ GeV)=$203-205$, 204 MeV.
These results are somewhat smaller than the ones obtained in the
nontruncation approach (\ref{res1}) but still larger than the ones
obtained in \cite{ChDo,JaMu}. The reason for this is that, as explained
in Sec.4,
the leading term on the theoretical side of the sum rule decreases
by about 6\% when going from $1/L^2$ to $1/L^3$. This is partly
compensated by a similar decrease in the contribution of the perturbative
continuum when truncated to the same order in $1/L$, such that the
final result for $m_s$ is smaller than in the untruncated approach.

Finally we should include the errors induced by the variation of the 
parameters (masses and widths) of the resonances and of the normalization
factor $d(s_{th})$. The former give an additional error of about $\pm$14
MeV on the value of $\hat m_s$ and of $\pm$15 MeV on $\bar m_s$(1 GeV).
The latter induces an error of about $\pm$(11-12) MeV on both mass parameters.

Adding all these errors in quadrature we obtain from (\ref{res1}) our 
final result
\bea\label{result}
\bar m_s(1 \mbox{ GeV}) = 205.5 \pm 19.1 \mbox{ MeV}\,.
\eea

This value lies on the high side of the existing QCD sum rule calculations
of $m_s$ \cite{Nar2,ChDo,JaMu}, coming closest to the recent
result obtained to three-loop order in \cite{Nar2}, of $196.7\pm 29.1$ MeV. 
The comparatively low results obtained in \cite{ChDo,JaMu} were in good
agreement with the lattice \cite{lattice} results for the strange quark mass.
With our new value the disagreement between the two is back in place.
In \cite{lattice} the value $\bar m_s(2 GeV)=128\pm 18$ MeV was obtained
 which gives $\bar m_s(1 GeV)=172\pm 24$ MeV. Recently, a
new lattice calculation has appeared \cite{lattice2} with lower
results: in the quenched approximation $\bar m_s$(2 GeV)=90$\pm 20$ MeV,
corresponding to $\bar m_s$(1 GeV)=$121 \pm 27$ MeV, and for $n_f=2$
an even lower value $\bar m_s$(2 GeV)=$70\pm 15$ MeV, respectively
$\bar m_s$(1 GeV)=$94\pm 20$ MeV.
Conceivable explanations for this discrepancy are a) significant systematic
errors in the parametrization of the hadronic density and b) large
contributions of direct instantons to the correlator of scalar currents
\cite{inst0,inst1,inst2}.
In our case, we consider b) to be little probable given the large scales
$M^2=2-9$ GeV$^2$ at which our determination is performed (for an
explicit calculation in the pseudoscalar current case see \cite{inst1}).
Thus further progress in improving the accuracy of the strange quark
mass determination using the methods of the present paper can only come
from a better knowledge of the hadronic density function. 
With the advent of a $\tau$-charm factory it should be
possible to directly measure it in the future in semileptonic $\tau$ decays.

\section*{Acknowledgements}
D.P. is supported  by a grant from the Ministry of Science and Arts of
Israel. He acknowledges the hospitality
 of the Theory Group of the Institute of Physics, Mainz,
during the final phase of this work.

\newpage
\section*{Figure captions}
\vspace*{1cm}
{\bf Fig.1} Dependence of the invariant strange quark mass $\hat m_s$
on the Borel parameter $M^2$ and on the continuum threshold $s_0$ (the
three lower curves). The upper three curves show the running mass at
the scale 1 GeV for the same values of the parameters.
($\Lambda_{QCD}=380$ MeV)\\[0.5cm]

\noindent
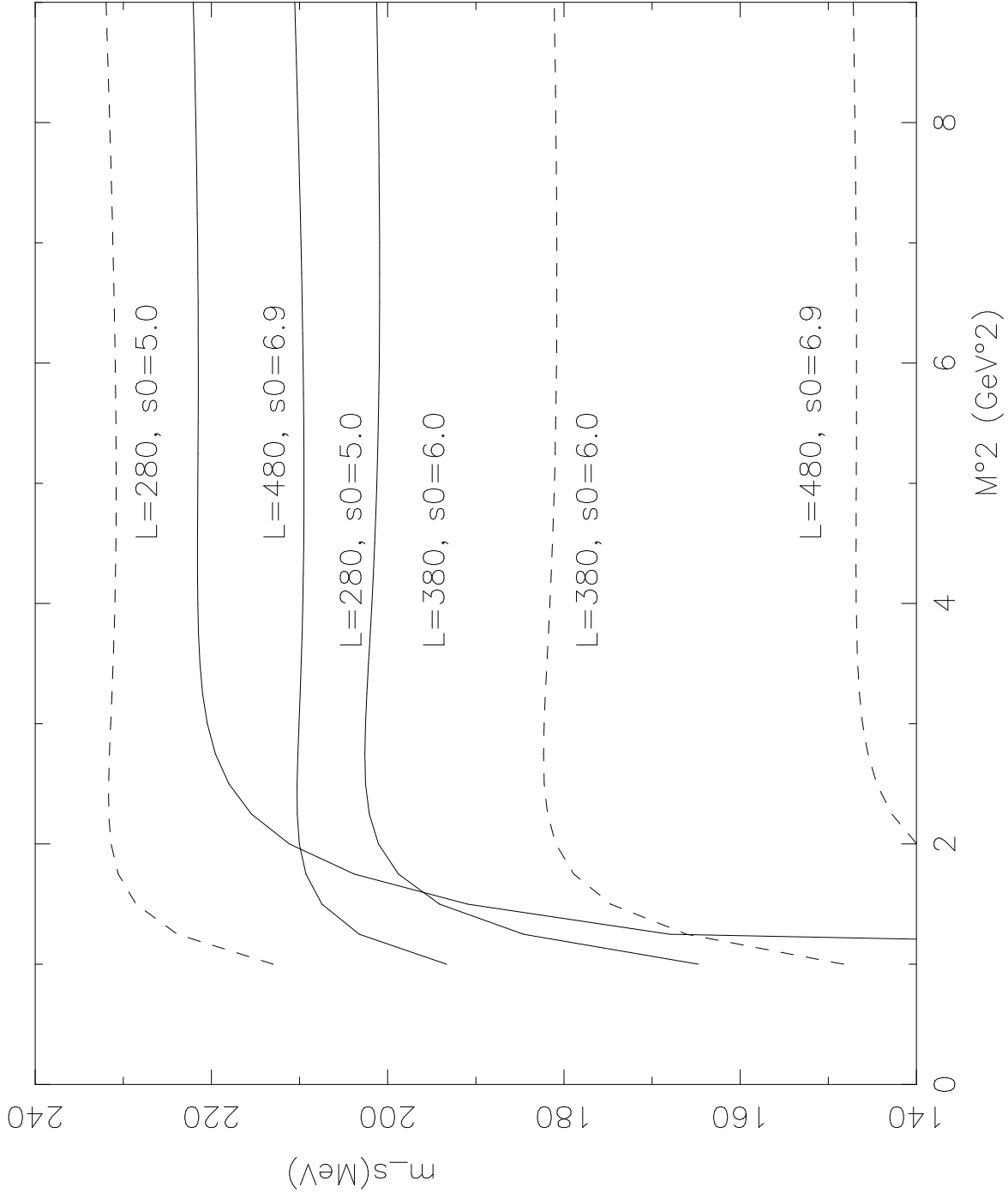
{\bf Fig.2} Dependence of the results on the value of the QCD scale
$\Lambda_{QCD}$. The continuous lines are the results for the running mass 
at the scale 1 GeV and the dotted lines show the invariant mass $\hat m_s$.

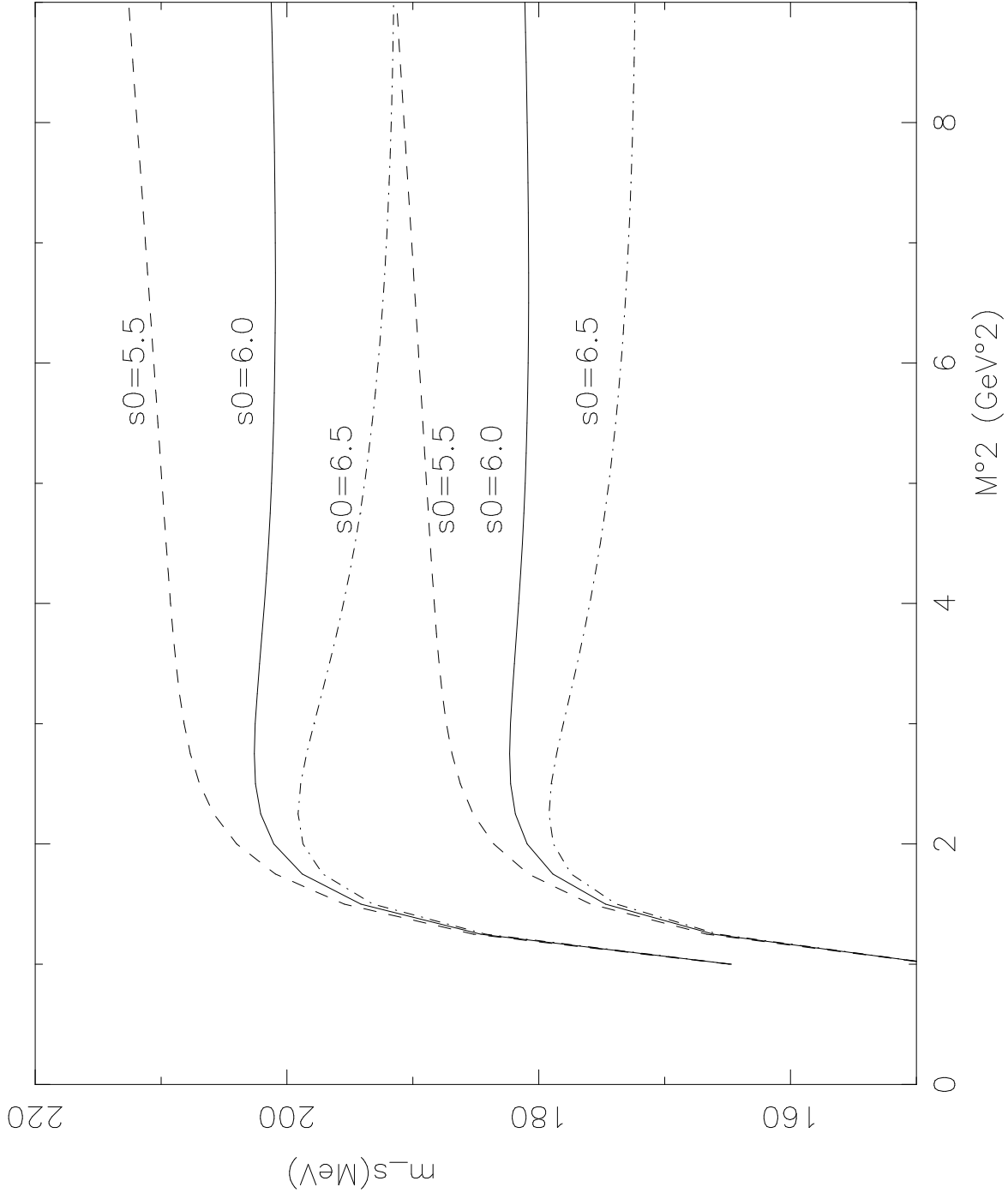
\end{document}